\documentclass[symmetry,article,submit,pdflatex,moreauthors]{mdpi} 

\firstpage{1} 
\pubvolume{0}
\issuenum{0}
\articlenumber{0}
\pubyear{2025}
\copyrightyear{2025}
\datereceived{ } 
\dateaccepted{ } 
\datepublished{ } 
\hreflink{https://doi.org/} 

\usepackage{subfigure}
\usepackage{slashed}
\usepackage{bm}
\usepackage[utf8]{inputenc}
\usepackage{newtxtext}
\usepackage[libertine]{newtxmath}
\usepackage[main=english]{babel}
\usepackage{microtype}
\usepackage{blindtext}

\usepackage{xcolor}
\usepackage[normalem]{ulem}
\definecolor{heavyred}{rgb}{0.750,0.000,0.000}

\newcommand{\vect}[1]{\boldsymbol{#1}}

\newcommand{\fslash}[1]{\mbox{$\!\not\!#1$}}
\newcommand{\be}{\begin{equation}}
\newcommand{\ee}{\end{equation}}

\def\emdash{\nobreak\hspace{0.1em}\textemdash\nobreak\hspace{0.1em}}

\graphicspath{{figures/},{gallery/}}
\def\emdash{\hspace*{1pt}---\hspace*{1pt}}

\Title{Effects of quark core sizes of baryons in neutron star matter}

\Author{Wolfgang Bentz $^{1,*}$ and Ian C. Clo\"et $^{2}$}

\AuthorNames{Firstname Lastname, Firstname Lastname and Firstname Lastname}

\AuthorCitation{Bentz, W.; Cloët, I.C.}
\TitleCitation{Effects of Quark Core Sizes of Baryons in Neutron Star Matter}

\address{$^{1}$ \quad Department of Physics, School of Science, Tokai University, 4-1-1 Kitakaname, 
  Hiratsuka-shi, Kanagawa 259-1292, Japan; bentz@tokai.ac.jp\\
$^{2}$ \quad Physics Division, Argonne National Laboratory, Argonne, Illinois 60438, USA; icloet@anl.gov}

\corres{Correspondence: bentz@tokai.ac.jp}

\abstract{We describe the quark substructure of hadrons and the equation of state of high density neutron star matter by using the Nambu--Jona-Lasinio (NJL) model, which is an effective quark theory based on QCD. The interaction between quarks fully respects the chiral and flavor symmetries. Guided by the success of various low energy theorems, we assume that the explicit breaking of these symmetries occurs only via the current quark masses, and all other symmetry breakings are of dynamical nature. In order to take into account the effects of the finite quark core sizes of the baryons on the equation of state, we make use of an excluded volume framework which respects thermodynamic consistency. The effects generated by the swelling quark cores generally act repulsively and lead to an increase of the pressure with increasing baryon density. On the other hand, in neutron star matter they also lead to a decrease of the density window where hyperons appear, because it becomes energetically more favorable to convert the faster moving nucleons into hyperons. Our quantitative analysis shows that the net effect of the excluded volume  is too small to solve the long standing ``hyperon puzzle'', which is posed by the large observed masses of neutron stars. Thus the puzzle persists in a relativistic effective quark theory which takes into account the short range repulsion between baryons caused by their finite and swelling quark core sizes in a phenomenological way.}

\keyword{Quark model; hyperons in nuclear matter; neutron stars} 

\begin{document}

\section{Introduction\label{sec:intro}}
The properties of strongly interacting hadronic systems like nucleons, nuclei, nuclear matter, and neutron stars, reflect the basic quark substructure of their constituents. At very high densities but low temperatures, as in the interior of neutron stars, new states of matter such as color superconducting quark matter~\cite{Alford:1997zt} or various kinds of coexisting phases of hadrons and quarks~\cite{Heiselberg:1999mq} are expected to play important roles, and to understand their structures is a fascinating field of current research. Besides the familiar blocks of nuclear systems\emdash protons and neutrons made of up ($u$) and down ($d$) quarks\emdash baryons and quarks with strangeness are receiving much attention experimentally as well as theoretically. The current $s$ quark mass ($m_s$) is still small enough to assume the full $SU(3)_L \otimes SU(3)_R$ symmetry of QCD for the effective hadronic interactions from intermediate to high energies. Thus $m_s$ can be viewed as the principal source of explicit chiral and flavor symmetry breakings; all other symmetry breaking patterns, including the more familiar vector $U_V(1)$~\cite{Weinberg:1986cq} and axial vector
$U_A(1)$~ \cite{tHooft:1976snw} cases, must be of dynamical origin. This way of thinking, which is supported by phenomenological and well as theoretical considerations~\cite{Gell-Mann:1962yej,Okubo:1961jc,Pagels:1974se}, puts strong constraints
on the hadronic interactions involving strangeness. Any theoretical model which aims to describe hyperons
in nuclei and nuclear matter, or strange quarks in hadrons and highly compressed neutron star matter, has to
comply with these constraints.

An important test stone for theoretical models was provided by the observation of heavy neutron stars with
about 2 solar masses~\cite{Demorest:2010bx,Antoniadis:2013pzd,Riley:2021pdl,Fonseca:2021wxt}, and more recently
up to 2.35 solar masses~\cite{Romani:2022jhd}. (For a recent analysis of independent data from the
NICER collaboration, including star radii, see Ref.~\cite{Brandes:2023bob}.)
Because the onset of hyperons usually leads to a softening
of the equation of state (EOS) of neutron star matter~\cite{Glendenning:1997wn}, it is still a major difficulty for
many models to reproduce such heavy stars. This problem, which is commonly called the ``hyperon puzzle''~\cite{Bombaci:2016xzl}, has triggered an extensive amount of research, see for example Ref.~\cite{Burgio:2021vgk}. Here we can mention only a few more recent examples:
Lattice QCD calculations, which have been very useful to study the hyperon-nucleon interactions~\cite{Beane:2012ey,Inoue:2018axd}, begin to provide new bounds on the pressure and the maximum mass
of neutron stars~\cite{Abbott:2024vhj,Moore:2023glb}. Nuclear density functional theory, based on
available empirical information on the interaction between hyperons~\cite{Choi:2023qtk}, relativistic
mean field theories with novel momentum-dependent interactions~\cite{Chorozidou:2024gyy}, and 
studies of three-body forces involving hyperons~\cite{Tamura:2022sme,Jinno:2023xjr}, are making
important contributions to understand the stiffening of the EOS of neutron star matter.
On the other hand, quark matter cores may exist in massive neutron stars under certain stability conditions~\cite{EslamPanah:2018rfe}, and their evidence
has been supported by model independent analyses~\cite{Annala:2019puf}.
In fact, detailed model calculations have shown that   
phase transitions to normal~\cite{Ferreira:2020evu} or color superconducting quark matter
~\cite{Contrera:2022tqh,Tanimoto:2019tsl} may be  
strong candidates to describe heavy neutron stars, if a vector-type repulsion
between the quarks in the core of the star is also taken into account.

Quarkyonic matter, which was originally motivated by the $1/N_c$ expansion~\cite{McLerran:2007qj,Jeong:2019lhv} and hadron-quark continuity~\cite{Baym:2017whm,Fukushima:2020cmk},
is receiving more and more interest recently because it may provide
a solution of the hyperon puzzle based on the Pauli blocking of $d$ quarks in neutron star
matter before the onset of hyperons~\cite{Fujimoto:2024doc,Kojo:2024ejq,Xia:2024wpz}. This is particularly
interesting because the role of short range repulsions arising from the Pauli principle on the quark level
in systems with strangeness
has also been suggested by recent group theoretical methods~\cite{Oka:2023hdc} and analysis of scattering
experiments~\cite{J-PARCE40:2022nvq}. A more phenomenological method to include short range repulsions
is based on the excluded volume effects (EVE) of the Van der Waals type, as used in previous studies of relativistic
heavy ion collisions~\cite{Rischke:1991ke,Yen:1997rv}. This method, which assumes that the finite sizes of quark cores in
baryons give rise to important short range repulsions, was used recently in Ref.~\cite{Leong:2023lmw}
in the framework of the quark-meson-coupling (QMC) model~\cite{Guichon:1987jp,Saito:2005rv},
where it was claimed that those effects are
strong enough to sustain heavy neutron stars. A closer investigation of this point
in a framework which respects the basic symmetries of QCD is the principal aim of our present investigation.

Nowadays it is well established that the Nambu--Jona-Lasinio (NJL) model~\cite{Nambu:1961tp,Nambu:1961fr,Vogl:1991qt,Hatsuda:1994pi}
is one of the most useful effective quark theories of QCD. A fully relativistic Faddeev
approach~\cite{Ishii:1995bu,Ishii:2000zy} has opened the
door to describe baryons in this model, and the inclusion of confinement effects in a
phenomenological way via an infrared cut-off~\cite{Bentz:2001vc} was the starting point for many
applications to nuclear systems~\cite{Cloet:2006bq,Cloet:2012td,Cloet:2015tha}. Despite quite
different model assumptions, the QMC model and the NJL model share many points in common,
in the present context most notably the important role of the scalar polarizability of the
nucleon to describe saturation and induced many-body forces~\cite{Birse:1994eg,Wallace:1994ux},
and the enhancement of spin-spin correlations between light quarks in
the nuclear medium which leads to an increase of the $\Sigma-\Lambda$ mass difference in nuclear matter~\cite{Guichon:2008zz,Noro:2023vkx}. In view of the results presented in Ref.~\cite{Leong:2023lmw} for the
QMC model, it is therefore of interest to see whether the NJL model can describe heavy neutron stars
in a similar way without invoking a phase transition to quark matter.

The purposes of our present work can therefore be summarized as follows: First, we wish to base the description
of excluded volumes on the actual quark core sizes, calculated consistently in the NJL model for each
baryon and each density separately, without introducing any new free parameters. It is well known that by assuming a common
quark core radius for all baryons in the system, a thermodynamically consistent description can be
achieved~\cite{Panda:2002iu} which satisfies the requirements of causality up to very high densities.
In view of our successful NJL model description of the properties of free octet baryons~\cite{Carrillo-Serrano:2016igi},
like their sizes, magnetic moments, and electromagnetic form factors, it is desirable to describe the
radii of in-medium quark cores and their effects to restrict the volumes available for the Fermi motion of
baryons in a consistent way. Because EVE reflect the short range repulsion
between nucleons in a phenomenological way, they will certainly stiffen the nuclear matter EOS.
However, it is well known that if this stiffening affects only
baryons in motion, i.e. mainly the neutrons, it becomes easier for them to reach
the hyperon thresholds, which leads to the unwanted effect of lowering the onset of hyperons~\cite{Baldo:1999rq}.
Therefore it is important to take into account also the effects of
finite\emdash actually swelling\emdash quark core radii of hyperons before their onset. Therefore 
we would like to consistently take into account the swelling of the quark cores of all in-medium baryons as function of the baryon density.\footnote{For the
nucleons, a moderate swelling is known to be important to describe the EMC effect~\cite{Cloet:2006bq}, the Coulomb sum rule~\cite{Cloet:2015tha}, and the reduction of the in-medium ratio of electric to magnetic nucleon form factors~\cite{Lu:1997mu}.}

Second, we wish to examine the effects of this extended description of EVE on the
EOS of nuclear matter, neutron star matter, and the properties of neutron stars in the framework of the mean field approximation to the NJL model.
The EOS and all numerical results of our present paper agree with our earlier work~\cite{Noro:2023vkx} using the standard
4-fermi interactions in the NJL model~\cite{Noro:2023vkx}, as soon as we switch off the EVE which arise from the finite quark core sizes. The basic physical picture is therefore the same, i.e., it can be visualized by composite baryons moving
in self consistent scalar and vector mean fields on the background of the constituent quark vacuum.
Using concepts of the Fermi liquid theory~\cite{Negele:1988aa,Shankar:1993pf}, it has been shown that this picture can be translated into a model of
baryons interacting via the exchange of composite neutral scalar and vector mesons~\cite{Bentz:2001vc,Noro:2023vkx}. In other words,
the overlap of the meson clouds of the baryons leads to the meson exchange interactions,
and it is natural to identify the distance at which the relative wave function of interacting baryons is
strongly suppressed with the quark core radii, i.e., the sizes of the baryons without their meson clouds.
Because we will calculate the quark core radii in our model without assuming any new parameters, we can investigate
their effects on the EOS in a parameter-free way. Because our previous NJL calculations~\cite{Noro:2023vkx} have
shown that 6-fermi and 8-fermi interactions\emdash with coupling constants chosen so as not to spoil the saturation
properties of normal nuclear matter\emdash cannot solve the hyperon puzzle, we limit ourselves to the standard
4-fermi interactions in this work.  

The outline of the paper is as follows: Sec.~\ref{sec:model} discusses our extended EOS in the NJL model, and demonstrates that thermodynamic consistency can be achieved in the case where each member of the baryon octet has its independent quark core size. Sec.~\ref{sec:results} presents
our results for the EOS of nuclear matter, neutron star matter, and the properties of neutron stars,
with particular emphasis on the EVE on the onset of hyperons, the role of the in-medium swellings of
baryons, and the resulting volume fractions occupied by the quark cores at very high baryon densities. Sec.~\ref{sec:summary} gives a summary of our results.

\section{NJL model and excluded volume effects\label{sec:model}}
Our work is based on the following three-flavor NJL Lagrangian with 4-fermi interactions in the $\bar{q}q$ channels~\cite{Vogl:1991qt,Hatsuda:1994pi}:
\begin{align}
\cal{L} &= \widebar{q} \left(i \fslash{\partial} - \hat{m} \right) q
+ G_{\pi} \left[ \left(\widebar{q} \lambda_a q\right)^{2} - \left(\widebar{q} \lambda_a \gamma_{5}  q\right)^{2} \right]  
- G_{v} \left[ \left(\widebar{q} \lambda_a \gamma^{\mu} q \right)^{2} 
 + \left(\widebar{q} \lambda_a \gamma^{\mu} \gamma_5 q \right)^{2} \right] \,,  
\label{eq:lagrangian}
\end{align}
where $q = (q_u, q_d, q_s)$ is the quark field, $\hat{m}$ the current quark mass matrix with diagonal elements $(m_u, m_d, m_s)$, and $\lambda_a$ ($a=0,1,2, \dots 8$) are the Gell-Mann flavor matrices plus $\lambda_0 = \sqrt{\frac{2}{3}} \vect{1}$. The 4-fermi coupling constants in the scalar--pseudoscalar and the vector--axial-vector channels are denoted by $G_{\pi}$ and $G_v$, respectively.
The interaction Lagrangian in Eq.~(\ref{eq:lagrangian}) has the  $SU(3)_L \otimes SU(3)_R \otimes U(1)_V \otimes U(1)_A$
symmetry of QCD, which contains the familiar flavor $SU(3)$ as a subgroup. 
In our previous work~\cite{Noro:2023vkx} we have shown that the 6-fermi (determinant) interaction~\cite{tHooft:1976snw}, which breaks the
$U(1)_A$ symmetry, and chiral invariant 8-fermi interactions~\cite{Osipov:2006ns} have only moderate effects on the EOS of neutron star matter if their strengths are restricted so as not to spoil the saturation properties of isospin symmetric nuclear matter.
Therefore, in this work we will consider only the 4-fermi interactions for simplicity.  
Following the successes of various low energy theorems and mass formulas, we will assume that current quark masses are the only
sources of explicit breaking of the flavor and chiral symmetries, and all other symmetry breakings are of dynamical nature.

In order to construct the octet baryons as quark-diquark bound states, we also need the interaction Lagrangian in the $qq$ channels with the same symmetries. This Lagrangian, which involves the 4-fermi coupling constants $G_S$ and $G_A$ in the scalar and axial-vector $qq$ channels, respectively, together with details of the quark-diquark description of octet baryons based on the Faddeev framework, were presented in our previous work (Appendix A of Ref.~\cite{Noro:2023vkx}),
and will not repeated here. In the vacuum isospin symmetry is assumed to be intact, i.e., we use $m_u = m_d \equiv m$ throughout this work. In neutron star matter, the isospin asymmetry will be fully taken into account by solving the bound state equations with independent
constituent quark masses $M_u$, $M_d$, and $M_s$ defined by Eq.~(\ref{eq:quarkmass}) below.

In order to construct the EOS we will use the mean field approximation, which can be visualized
by composite baryons moving in self consistent scalar and vector fields on the background of the constituent quark vacuum.
We will take into account three scalar fields $\sigma_{\alpha}$ and the time components of three
4-vector fields $\omega_{\alpha} \equiv \omega_{\alpha}^0$, where $\alpha= u, d, s$. Their definitions are as follows:
\begin{align}
\sigma_{\alpha} &= 4 G_{\pi} \langle \widebar{q}_{\alpha} \, q_{\alpha} \rangle \,, \,\,\,\,\,\,\,\,\,\,\,\,\,\,\,
\omega_{\alpha} = 4 G_v \langle \widebar{q}_{\alpha}\gamma^0 q_{\alpha} \rangle \,,
\label{eq:fields}
\end{align}
where $\langle \ldots \rangle$ denotes the expectation value in the ground state of the medium under consideration
(vacuum, nuclear matter, or neutron star matter). 
The scalar fields lead to spontaneous breaking of the chiral symmetry and give rise to the effective quark masses
\begin{align}
M_{\alpha} = m_{\alpha} - \sigma_{\alpha} \,.
\label{eq:quarkmass}
\end{align}
The vector fields lead to shifts of the energies of the baryons in the system. As a result, the energy of a baryon with flavor $b$ and 3-momentum $\vect{k}$ is obtained from the pole of the quark-diquark equation in the variable
$k_0$ as\footnote{Here and in the following, a summation over multiple flavor indices ($\alpha, \beta$ for quarks, $b, b'$ for octet baryons, $\ell$ for leptons, and $i$ for both baryons and leptons) in a product, including squares like $\omega_{\alpha}^2$, is implied if those indices appear on
the same side of an equation. (As usual, the same convention is used for the Lorentz indices $\mu, \nu$.) The Fermi momentum of particle $i$ will be denoted as $p_i$.}  
\begin{align}
\varepsilon_b(k) = \sqrt{\vect{k}^2 + M_b^2}  + n_{\alpha/b} \, \omega_{\alpha} 
\equiv E_b(k) +  n_{\alpha/b} \, \omega_{\alpha} \,,
\label{eq:baryonenergy}
\end{align}
where $n_{\alpha/b}$ is the number of quarks with flavor $\alpha$ in the baryon $b$ with mass $M_b$.

The mean field approximation is implemented into the Lagrangian~(\ref{eq:lagrangian}) in the standard way by decomposing the various quark bilinears into classical (c-number) parts and quantum (normal ordered) parts. We will assume that the only non-vanishing classical parts are the mean fields given in Eq.~(\ref{eq:fields}).  The normal ordered parts, together with the $q q$ interaction parts given in Eq.~(A1) of Ref.~\cite{Noro:2023vkx}, are used to calculate bound state masses of pseudoscalar mesons and octet baryons, as well as the pion decay constant.

\subsection{Mean field approximation without EVE\label{sec:mfa}}
In our model defined above, the energy density of neutron star matter has the form\footnote{Here we attach a superscript $(0)$ to those quantities which, in the next subsection,  will be considered as effective ones
  if their argument is the set of effective baryon densities ${\tilde \rho}$, defined by Eq.~(\ref{eq:bdens}) of the next subsection.}
\begin{align}
  {\cal E}^{(0)}(\rho) &= {\cal E}_B^{(0)}(\rho)
                   + {\cal E}_{\rm vac} - \frac{\omega_{\alpha}^2}{(8 G_v)} 
                   + {\cal E}_{\ell}  \,.
\label{eq:e1}
\end{align}
Here
\begin{align}
  {\cal E}_B^{(0)}(\rho) &= 2 \int \frac{{\rm d}^3k}{(2\pi)^3} \varepsilon_b(k) \, n_b(k, p_b),
  \label{eq:ebary}
\end{align}
is the energy density of the baryons, where $\rho \equiv \{\rho_{b_1}, \rho_{b_2}, \dots \}$ denotes a given set of octet baryon densities, $n_b(k, p_b) = \theta(p_b - k)$ are the Fermi distributions with Fermi momenta $p_b$, and the single baryon energies were given in Eq.~(\ref{eq:baryonenergy}). The total baryon density will be denoted as $\rho_B \equiv \sum_i \rho_{b_i}$. ${\cal E}_{\rm vac}$ in Eq.~(\ref{eq:e1}) is the (Mexican-hat shaped) contribution of the constituent quark vacuum to the energy density, expressed in its unregularized form by 
\begin{align}
  {\cal E}_{\rm vac}  = 6 i \int  \frac{{\rm d}^4 k}{(2 \pi)^4} \, 
\ln \frac{k^2 - M_{\alpha}^2}{k^2 - M_{\alpha 0}^2} 
+ \frac{\sigma_{\alpha}^2 - \sigma_{\alpha 0}^2}{8 G_{\pi}} \,,
  \label{eq:vacuum}
\end{align}
where a sum over the quark flavors is implied, and the sub-index $0$ refers to the vacuum with
zero baryon densities.
The label $\ell = e^-, \mu^-$ in Eq.~(\ref{eq:e1}) denotes the contributions from the
Fermi gas of leptons in chemical equilibrium with the baryons.

The stationary conditions for the vector fields $\frac{\partial {\cal E}^{(0)}}{\partial \omega_{\alpha}} = 0$ for fixed $\rho$ gives
$\omega_{\alpha} = 4 G_v \, \rho_{\alpha} = 4 G_v \, n_{\alpha/b} \, \rho_b$,
    and therefore the total contribution of the vector fields to the energy density becomes
$\frac{\omega_{\alpha}^2}{8 G_v} = 2 G_v \rho_{\alpha}^2$.
The corresponding requirement on the scalar fields, $\frac{\partial {\cal E}^{(0)}}{\partial \sigma_{\alpha}} = 0$, is equivalent to the
  in-medium gap equations. The energy density
is then stationary w.r.t. all mean fields, and the chemical potential of baryon $b$ becomes simply the Fermi energy $\varepsilon_b(k=p_b)$:
\begin{align}
  \mu^{(0)}_b(\rho) &\equiv \frac{{\rm d}\, {\cal E}^{(0)}(\rho)}{{\rm d} \rho_b} =
                  \frac{\partial\, {\cal E}^{(0)}(\rho)}{\partial \rho_b}  
                = E_b(p_b) + n_{\alpha/b} \, \omega_{\alpha} \,.
 \label{eq:mu}
\end{align}

Because the model is thermodynamically consistent, the pressure can be calculated either from the first law of thermodynamics
($P^{(0)}(\rho) + {\cal E}^{(0)}(\rho) = \rho_b \, \mu_b^{(0)}(\rho) + \rho_{\ell} \mu_{\ell}$) or by direct calculation.
The resulting expression is
\begin{align}
  P^{(0)}(\rho) &= P_B^{(0)}(\rho)  
    - {\cal E}_{\rm vac} + \frac{\omega_{\alpha}^2}{(8 G_v)} + P_{\ell} \,,
\label{eq:pr}
\end{align}
where the contribution from the baryons is given by
\begin{align}
  P_B^{(0)}(\rho) &= \rho_b \, \mu_b^{(0)}(\rho) - {\cal E}_B^{(0)}(\rho) =
    \frac{2}{3}  \int \frac{{\rm d}^3k}{(2\pi)^3} \frac{k^2}{E_b(k)} \, n_b(k, p_b) \,. 
    \label{eq:prb}
\end{align}

\subsection{EVE for baryons with different quark core sizes\label{sec:eve}}
In order to take into account the EVE, we define the effective baryon densities by 
$\tilde{\rho}_b = \frac{N_b}{V - V_{qc}}$, where $N_b$ are true baryon numbers in the volume $V$, and $V_{qc}$ is the volume
occupied by the quark cores of all baryons. 
These effective baryon densities are considered as functions of the true baryon densities $\rho_b = \frac{N_b}{V}$
according to the relation
\begin{align}
\tilde{\rho}_b(\rho) = \frac{\rho_b}{1 - v_{b'}\rho_{b'}} \equiv \frac{\rho_b}{1 - v \cdot \rho} \,,
\label{eq:bdens}
\end{align}
where $v_{b}$ is the volume of the quark core of baryon $b$, which will be defined more precisely in the next section,
and therefore $v_b \, \rho_b \equiv v \cdot \rho$
is the volume fraction ($V_{qc}/V$) occupied by the quark cores of all baryons.

Let us first consider the pressure. To
take into account the EVE , we follow previous works~\cite{Rischke:1991ke,Yen:1997rv,Panda:2002iu,Leong:2023lmw} and replace the explicit dependence on the baryon densities
$\rho = \{\rho_1, \rho_2, \dots \}$ in the last form of (\ref{eq:prb}) by the
effective baryon densities $\tilde{\rho} \equiv \{\tilde{\rho}_1, \tilde{\rho}_2, \dots \}$,
i.e. we use
\begin{align}
  P_B(\rho) = P_B^{(0)}(\tilde{\rho}) =
  \frac{2}{3}  \int \frac{{\rm d}^3k}{(2\pi)^3} \frac{k^2}{E_b(k)} \, n_b(k, \tilde{p}_b) 
  \label{eq:press}
\end{align}
with the effective Fermi momenta $\tilde{p}_b$ defined by $\tilde{\rho}_b = \tilde{p}_b^3/(3 \pi^2)$.  
The total pressure as a function of the true baryon densities then follows from Eq.~(\ref{eq:pr}) as
\begin{align}
P(\rho) = P_B^{(0)}(\tilde{\rho}) -  {\cal E}_{\rm vac} + \frac{\omega_{\alpha}^2}{(8 G_v)} + P_{\ell} \,.
\label{eq:presst}
\end{align}   

The effective baryon energy density is defined as the energy of the baryons per volume available for their motion, i.e.,
\begin{align}
  {\cal E}_B^{(0)}(\tilde{\rho}) \equiv \frac{{\cal E}_B(\rho)} {1 - v \cdot \rho}  \,.    \nonumber
\end{align}
Therefore the baryon energy density including the EVE is expressed as
${\cal E}_B(\rho)=\left(1 - v \cdot \rho\right) \, {\cal E}^{(0)}_B(\tilde{\rho})$,
where ${\cal E}^{(0)}_B(\tilde{\rho})$ is given by the original NJL model expression (\ref{eq:ebary}) after the
replacement $p_b \rightarrow \tilde{p}_b$:
\begin{align}
  {\cal E}^{(0)}_B(\tilde{\rho}) &= 2 \int \frac{{\rm d}^3k}{(2\pi)^3} \varepsilon_b(k) \, n_b(k, \tilde{p}_b) \,.
  \label{eq:ebary1}
\end{align}
We therefore arrive at the following expression for the total energy density including the EVE:\footnote{We note that the factor $\left(1 - v \cdot \rho \right)$ in Eq.~(\ref{eq:ed}) does not mean that only the contributions from outside the quark cores are taken into account, as can be seen for example by the correct zero density limit ${\cal E}(\rho) \rightarrow M_b \, \rho_b$.}
\begin{align}
  {\cal E}(\rho) = \left(1 - v \cdot \rho \right) \, {\cal E}^{(0)}_B(\tilde{\rho})
   +  {\cal E}_{\rm vac} - \frac{\omega_{\alpha}^2}{(8 G_v)} + {\cal E}_{\ell} \,.
\label{eq:ed}
\end{align}
The mean fields including the EVE are determined for fixed true baryon densities
by the stationarity conditions applied to (\ref{eq:ed}). 
By using the spectrum (\ref{eq:baryonenergy}), for the vector fields this condition leads to
  \begin{align}
    \frac{\partial {\cal E}(\rho)}{\partial \omega_{\alpha}} = (1 - v \cdot \rho) \, n_{\alpha/b} \, \tilde{\rho}_b -
    \frac{\omega_{\alpha}}{4 G_v} = n_{\alpha/b} \, \rho_b - \frac{\omega_{\alpha}}{4 G_v} = 0 \,,
    \label{eq:stat}
  \end{align}
  which shows that the vector fields are unchanged by the EVE:
  $\omega_{\alpha} = 4 G_v \, n_{\alpha/b} \, \rho_b = 4 G_v \rho_{\alpha}$.
  The total vector contribution to the energy density (\ref{eq:ed}) is then the same as in the previous subsection, namely
  $\frac{\omega_{\alpha}^2}{8 G_v} = 2 G_v \rho_{\alpha}^2$.
  For the scalar fields, the gap equations
\begin{align}
    \frac{\partial\, {\cal E}(\rho)}{\partial \sigma_{\alpha}}=0
    \label{eq:gap}
\end{align}
have to be solved numerically to obtain the quark and baryon masses as functions of the true baryon densities.   Because from Eqs.~(\ref{eq:stat}) and (\ref{eq:gap}) the energy density is stationary w.r.t. all mean fields,   the baryon chemical potential can be obtained from   $\mu_b(\rho) \equiv {\rm d} {\cal E}(\rho)/{\rm d} \rho_b =  {\partial} {\cal E}(\rho)/\partial \rho_b$, i.e., only the explicit dependence on $\rho$, which resides in the first term of (\ref{eq:ed}), needs to be taken into account. A simple calculation, using the relation $\mu_b^{(0)}(\tilde{\rho})= \partial {\cal E}^{(0)}(\tilde{\rho})/(\partial \tilde{\rho}_b)$ and the thermodynamic consistency relation of the original NJL model  (${\cal E}^{(0)}_B(\tilde{\rho}) + P^{(0)}_B(\tilde{\rho}) = \tilde{\rho}_b \,   \mu_b^{(0)}(\tilde{\rho})$), gives
\begin{align}
  \mu_b(\rho) &= - v_b \, {\cal E}^{(0)}_B(\tilde{\rho}) + \left(1 - v \cdot \rho \right)
  {\mu}^{(0)}_{b'}(\tilde{\rho}) \left( \frac{\delta_{b', b}}{1 - v \cdot \rho}
                + \frac{\rho_{b'} v_{b}}{(1 - v \cdot \rho)^2} \right)  \nonumber \\ 
        & = {\mu}^{(0)}_b(\tilde{\rho})  - v_b \left( {\cal E}^{(0)}_B(\tilde{\rho}) -
                {\mu}^{(0)}_{b'}(\tilde{\rho}) \, \tilde{\rho}_{b'} \right)  
          = \mu_b^{(0)}(\tilde{\rho}) + v_b \, P^{(0)}_B({\tilde \rho}) \nonumber \\
     &=  E_b(\tilde{p}_b) + n_{\alpha/b} \, \omega_{\alpha} + v_b \, P_B(\rho) \,,
\label{eq:mub3}
\end{align}
where in the last step we used the form of $\mu_b^{(0)}(\tilde{\rho})$ as given by the last form in Eq.~(\ref{eq:mu}) after the replacement $p_b \rightarrow \tilde{p}_b$, and the relation $P_B(\rho) = P_B^{(0)}(\tilde{\rho})$ according to Eq.~(\ref{eq:press}). From Eq.~(\ref{eq:mub3}) we see that the EVE led to an increase of the Fermi momenta of baryons, and to a volume term in the chemical potentials.
  
We can now confirm the thermodynamic consistency of the approach, that is the first law of thermodynamics,
\begin{align}
{\cal E}(\rho) + P(\rho) = \mu_b(\rho) \, \rho_b + \mu_{\ell} \, \rho_{\ell}
\label{rel2}
\end{align}
as follows: For the leptonic contributions it is obvious, and for the other contributions we use
the form (\ref{eq:presst}) for $P(\rho)$ and (\ref{eq:ed}) for ${\cal E}(\rho)$, as well as Eq.~(\ref{eq:mub3}) and the aforementioned thermodynamic consistency relation of the original NJL model for the baryon contributions.
Leaving out the leptonic contributions for simplicity, we get 
\begin{align}
  {\cal E}(\rho) + P(\rho) &= {\cal E}^{(0)}_B(\tilde{\rho}) + P^{(0)}_B(\tilde{\rho}) -
          \left(v \cdot \rho \right) {\cal E}^{(0)}_B(\tilde{\rho}) 
        = \mu^{(0)}_b(\tilde{\rho}) \, \tilde{\rho}_b  - \left(v \cdot \rho \right)
        {\cal E}^{(0)}_B(\tilde{\rho}) \nonumber \\ 
   &= \left(\mu_b(\rho) - v_b P^{(0)}_B(\tilde{\rho}) \right) \frac{\rho_b}{1 - v \cdot \rho}
     - \left(v \cdot \rho \right) \left( \mu^{(0)}_b(\tilde{\rho}) \, \tilde{\rho}_b - P^{(0)}_B(\tilde{\rho})
     \right)  \nonumber \\ 
   &= \mu_b(\rho) \, \rho_b \left(1 + \frac{v \cdot \rho}{1 - v \cdot \rho} \right) 
     - \left(v \cdot \rho \right) \left(\mu_b(\rho) - v_b P^{(0)}_B(\tilde{\rho}) \right) \frac{\rho_b}{1 - v \cdot \rho}
     \nonumber \\
     &+ \left(v \cdot \rho \right) P^{(0)}_B(\tilde{\rho}) \left( 1 - \frac{1}{1 - v \cdot \rho} \right),                       
\end{align}
which becomes
\begin{align}
{\cal E}(\rho) + P(\rho)
&= \mu_b(\rho) \, \rho_b + \left(v \cdot\rho \right) P_B^{(0)}(\tilde{\rho}) \left( 1 + \frac{v \cdot \rho}{1 - v \cdot \rho} - \frac{1}{ 1 - v \cdot \rho} \right)
= \mu_b(\rho) \, \rho_b \,.
\label{check}                          
\end{align}
Compared to previous works on EVE in effective quark theories~\cite{Panda:2002iu,Leong:2023lmw}, the important new point here is that the thermodynamic consistency can be satisfied for the case where each baryon has its own quark core radius,  which can be calculated consistently in the same model framework, as will be explained in the next section. We finally note that the formulas given in this subsection can be used to derive the following simple and instructive relation for the change of the energy density w.r.t. the quark core volume of a baryon $b$:
\begin{align}
\frac{\partial\,{\cal E}}{\partial v_b} = \rho_b \, P_B(\rho) \,.   
\label{eq:dedb}
\end{align}

\section{Results\label{sec:results}}
In this section we present our results on EVE in symmetric nuclear matter, neutron star matter, and neutron stars. The EVE in our model arise from the finite rms radii of the baryon density distributions of the quark cores of octet baryons in the medium, which are calculated by using the quark-diquark model based on the Faddeev equation for each baryon separately, as will be explained below.  

\subsection{Model parameters and results for single baryons\label{sec:parameters}}

\begin{table}[tbp]
\addtolength{\extrarowheight}{2.2pt}
\centering
\caption{Values for the model parameters which are determined in the vacuum, single hadron, and nuclear matter sectors. The regularization parameters, constituent quark masses in the vacuum (sub-index $0$) and current quark masses are given in units of GeV, and the coupling constants in units of GeV$^{-2}$.}
\begin{tabular*}{\columnwidth}{@{\extracolsep{\fill}}cccccccc}
\hline\hline
$\Lambda_{\rm IR}$  & $\Lambda_{\rm UV}$  &  $G_{\pi}$  &  $G_v$  &  $M_0$  &  $M_{s0}$ &  $m$ & $m_s$ \\
\hline
0.240 & 0.645 & 19.04 & 6.03 & 0.40 & 0.562 & 0.016 & 0.273  \\
\hline\hline
\end{tabular*}
\label{tab:parameters}
\end{table}

The model parameters used in the present numerical calculations are the same as in our previous work~\cite{Noro:2023vkx}, and are summarized in Tab.~\ref{tab:parameters}. Besides the model parameters $G_{\pi}$, $G_v$, $m$ and $m_s$, which appear in the Lagrangian of Eq.~(\ref{eq:lagrangian}), other parameters necessary to define the model are      the infrared (IR) and ultraviolet (UV) cut-offs $\Lambda_{\mathrm{IR}}$ and $\Lambda_{\mathrm{UV}}$, which are used with the proper-time regularization scheme~\cite{Schwinger:1951nm,Hellstern:1997nv}, see also App.~C of Ref.~\cite{Noro:2023vkx}.  While the UV cut-off is necessary to give finite integrals, the IR cut-off eliminates unphysical decay thresholds of hadrons into quarks, thereby simulating one important aspect of confinement. It should be similar to $\Lambda_{\rm QCD}$, and because our results are rather insensitive to its precise value as long as $0.20 \, {\rm GeV} < \Lambda_{\rm IR} < 0.28 \, {\rm GeV}$ we  choose $\Lambda_{\rm IR} = 0.24\,$ GeV.\footnote{A recent study~\cite{Kneur:2011vi} of $\Lambda_{\rm QCD}$ in the $\overline{\rm MS}$ scheme for 2 active flavors gave $0.255^{+0.04}_{-0.015}$ GeV, which encompasses our assumed value.} The parameters   $\Lambda_{\rm UV}$, $m$, and $G_{\pi}$ are determined so as to give a constituent light quark mass in vacuum of $M_0 = 0.4\,$GeV via the gap equation (see Eqs.(\ref{eq:quarkmass}) and (\ref{eq:gap})), the pion decay constant $f_{\pi} = 0.93\,$GeV, and pion mass $m_{\pi} = 0.14\,$GeV via the standard Bethe-Salpeter equation in the pionic $\overline{q} q$ channel~\cite{Vogl:1991qt,Hatsuda:1994pi}. The strange quark mass $m_s$ is determined so that the gap equation gives a constituent $s$-quark mass in vacuum of $M_{s0} = 0.562$ GeV, which reproduces the observed mass of the $\Omega$ baryon $M_{\Omega} = 1.67$ GeV by using the quark-diquark bound state equations. The vector coupling $G_{v}$ is determined from the binding energy per-nucleon in symmetric nuclear matter ($E_B/A = -16 \,$ MeV) at the saturation density ($\rho_{B0} = 0.15\,$ fm$^{-3}$).\footnote{See Refs.~\cite{Horowitz:2020evx,Wang:2014qqa} for recent precision studies of $\rho_{B0}$ and $E_B/A$ at $\rho_B=\rho_{B0}$ of symmetric nuclear matter.} As will be seen in the next subsection, the EVE lead to a slight shift of the saturation point of symmetric nuclear matter which, however, is completely unimportant for neutron star matter at high densities, so we will keep the value of $G_v$ determined in our previous work~\cite{Noro:2023vkx}. As explained in Appendix A of Ref.~\cite{Noro:2023vkx}, the coupling constants in the scalar and axial vector $qq$ channels, $G_S$ and $G_A$, are fixed to the free nucleon and delta masses ($M_{N0} = 0.94$ GeV, $M_{\Delta 0} = 1.23$ GeV). Except for $M_{N0}$, the resulting free masses of octet baryons are then predictions of the model, and are summarized in Tab.~\ref{tab:octetmasses} together with the observed values.\footnote{The good agreement with the observed values, and in particular with the Gell-Mann Okubo octet mass relation~\cite{Gell-Mann:1962yej,Okubo:1961jc}  $M_{N0} + M_{\Xi0} = \frac{1}{2} \left(M_{\Sigma0} + 3 M_{\Lambda0}\right)$, may be taken as an a posteriori justification of our choice $M_0 = 0.4$ GeV.}

\begin{table}[tbp]
  \addtolength{\extrarowheight}{2.2pt}
    \centering
    \caption{Masses of octet baryons (in units of GeV up to 3 significant figures)
      calculated in the vacuum (sub-index $0$)
    by using the coupling constants $G_S = 8.76$ GeV$^{-2}$ and $G_A = 7.36$ GeV$^{-2}$ in the scalar
    and axial vector diquark channels fitted to the standard vacuum masses of the
    nucleon ($M_{N0}=0.940$ GeV) and the $\Delta$ baryon ($M_{\Delta} = 1.32$ GeV),
    in  comparison to the isospin-averaged observed values~\cite{ParticleDataGroup:2024cfk}.
}
    \begin{tabular*}{\columnwidth}{@{\extracolsep{\fill}}ccccc}
        \hline\hline
            & $M_{N0}$ & $M_{\Lambda 0}$ & $M_{\Sigma 0}$
        & $M_{\Xi 0}$ 
        \\ \hline
        calc. & 0.940 & 1.12 & 1.17 & 1.32 \\
        obs.  & 0.939 & 1.12 & 1.19 $\pm$ 0.02 & 1.32 $\pm$ 0.14 \\
        \hline\hline
    \end{tabular*}
    \label{tab:octetmasses}
\end{table}

\begin{figure}[t]
  \centering\includegraphics[width=0.9\columnwidth]{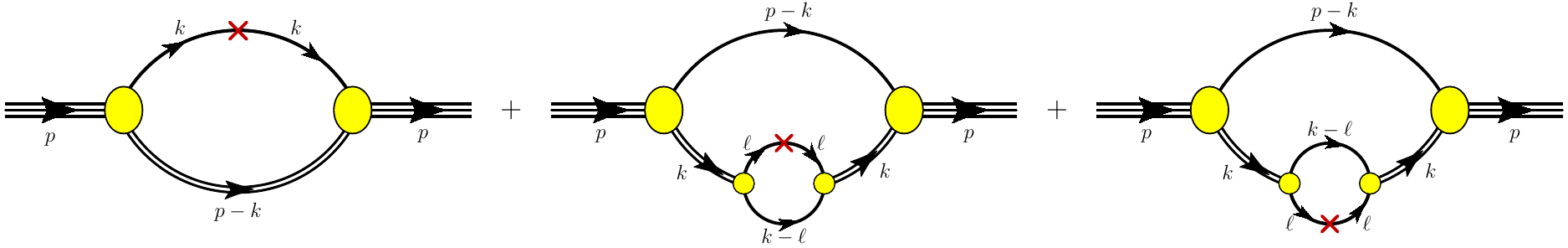}  
  \caption{Feynman diagrams representing the quark core contributions to the baryon current.
    The ovals (circles) are the vertex functions of the
    quark-diquark (quark-quark) bound state as given in
      Refs.\cite{Noro:2023vkx,Carrillo-Serrano:2014zta,Carrillo-Serrano:2016igi}, and the crosses stands for the external
    field which transfers a 4-momentum
    $q$ to the baryon and couples to the constituent quarks via the elementary vertex $\frac{1}{3} \gamma^{\mu}$.
    A single (double)
    solid line stands for the quark (diquark) propagator, and the external lines represent a given
    member of the baryon octet.}
  \label{fig:feynman}
\end{figure}

By using the solutions of the quark-diquark bound state equations as described in
Appendix A of Ref.~\cite{Noro:2023vkx},  we
can calculate the quark core (qc) contribution to the form factor corresponding to the distribution of baryon
charge from the Feynman diagrams
shown in Fig.~\ref{fig:feynman}, which represent the baryon current associated with ``bare'' constituent quarks, i.e.,
constituent quarks without their meson clouds. Each operator insertion in these diagrams
is therefore given by $\frac{1}{3} \gamma^{\mu}$, where the factor $\frac{1}{3}$ is the baryon number of a quark.

Similar to the more familiar electric form factor, the baryon current represented by Fig.~\ref{fig:feynman} can be parametrized by the Dirac-Pauli form factors
$F^{(qc)}_{1b}$ and $F^{(qc)}_{2b}$, which give the desired form factor as $G^{(qc)}_b(Q^2) = F^{(qc)}_{1b}(Q^2) - \frac{Q^2}{4 M_b^2} F^{(qc)}_{2b}(Q^2)$ and the associated rms radius $r_b^{(qc)} = \sqrt{ - 6 \, {\rm d}G^{(qc)}_b/{\rm d}Q^2|_{Q^2=0}}$. Our results for these
radii in the vacuum (zero baryon density), where isospin symmetry is assumed to be intact, are shown in the second
column of Tab.~\ref{tab:radii}. It is important to note that it is these baryon radii of quark cores without
meson cloud effects which are responsible for the EVE in our model, and which precisely agree with the notion of
``hard core radii'' of Ref.~\cite{Fukushima:2020cmk}.
The physical reason is simply that the overlap of the
meson clouds gives rise to the meson exchange interaction, and should not be counted as an excluded volume effect.
Indeed, by using familiar relations of the Fermi liquid theory, one can verify that our effective baryon-baryon interaction
takes into account the direct terms of neutral scalar and vector meson exchanges (see App.~\ref{app:interaction}).
They mediate the interactions between the baryons by generating the mean fields, and do not restrict the volume
available for their motion in any way. It is interesting to observe that our values of $r_b^{(qc)}$ are similar to the
radii used in previous works on EVE~\cite{Panda:2002iu,Leong:2023lmw}, where they were used as free parameters common to all
baryons.

\begin{table}[h]
\addtolength{\extrarowheight}{2.2pt}
    \centering
    \caption{Baryon radii of single (free) baryons (in units of fm up to 2 significant figures).
      The second column shows the
      quark core (qc) results, which are used to assess the EVE on the EOS in this
      work. The third and fourth columns, which are taken from our earlier calculations~\cite{Carrillo-Serrano:2016igi}
      in the same model, include also the effects of the vector meson (vm) and the pion ($\pi$) cloud for those cases where the baryon radii can be related to the electric radii, as explained in footnote 8.
      For those cases, the last column shows the observed values~\cite{ParticleDataGroup:2024cfk} where available.
      For the $\Xi$, where no observed value is available, we quote the results of a recent calculation in nonlocal chiral
      effective theory~\cite{Yang:2020rpi}.}
    \begin{tabular*}{\columnwidth}{@{\extracolsep{\fill}}ccccc}
        \hline\hline
        $b$           &   $r_b^{({\rm qc})}$   & $r_b^{({\rm qc+vm})}$     & $r_b^{({\rm qc+vm}+\pi)}$  &  $r_b^{\rm obs}$
        \\ \hline
      $N$          & $0.47$               & $0.78$                 & $0.79$   &   $0.77$  \\
      $\Sigma$     & $0.46$               & $0.74$                 & $0.86$   &   $0.78 \pm 0.10$ \\
      $\Xi$        & $0.44$               & $0.69$                 & $0.76$   &   $0.77 \pm 0.08$ \\
      $\Lambda$    & $0.46$               &                        &          &                   \\
        \hline\hline
    \end{tabular*}
    \label{tab:radii}
\end{table}

Although the meson cloud effects, a part of which is incorporated automatically in our effective interaction, should not be included in the definition of hard core radii,
they are of course very important to explain the physical sizes of single baryons. For the case of the baryon radii $r_b$,
the vector mesons are of particular importance, as one can expect from the vector meson dominance model, but also the pion cloud gives some contributions. For illustrative
purposes we therefore quote the results from Table VI of our previous work~\cite{Carrillo-Serrano:2016igi} on baryon octet form factors, which was
performed within the same model, in the third and fourth columns of Tab.~\ref{tab:radii}.
\footnote{For the nucleon, the relation between the baryon radius and the electric charge radii of protons and neutrons is
  $r_N^2 = \big<r_{Ep}^2\big> + \big<r_{En}^2\big>$. For the $\Sigma$ and $\Xi$ the corresponding relations are simply $r_{\Sigma}^2 = - \big<r_{E \Sigma^-}^2\big>$ and $r_{\Xi}^2 = -\left<r_{E \Xi^-}^2\right>$, because $\Sigma^-$ and $\Xi^-$ consist only of $d$ and $s$ quarks with the same electric charge of $-1/3$. No such relation exists for the $\Lambda$.} (For more detailed discussions for the case of nucleons, see Ref.~\cite{Cloet:2014rja}.)

Because in the following discussions the term ``baryon radius'' always refers to the quark core radius $r_b^{({\rm qc})}$,
we return to the simplified notations of the previous section, i.e., we use $r_b$ for the radius and
$v_b=\frac{4\pi}{3} r_b^3$ for volume. The values of $r_b$ in the vacuum, given in the second column
of Tab.~\ref{tab:radii}, will be denoted by $r_{b0}$.

\begin{figure}[tbp]
\subfigure{\centering\includegraphics[width=0.48\columnwidth]{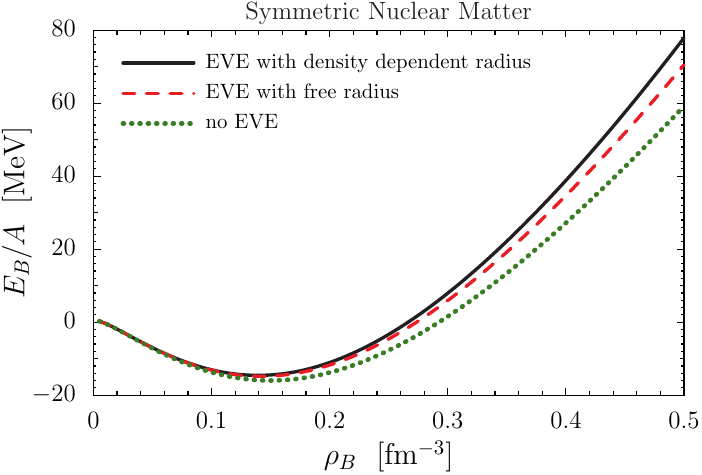}} \hfill
  \subfigure{\centering\includegraphics[width=0.48\columnwidth]{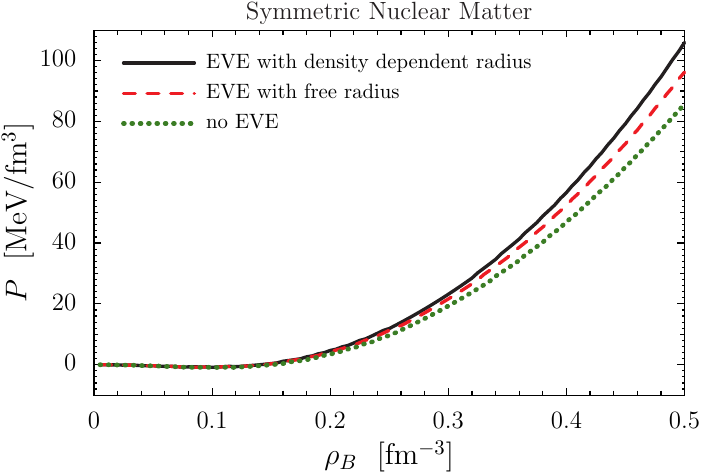}}
  \caption{Binding energy per nucleon ({\it left panel}) and pressure ({\it right panel}) in
    symmetric nuclear matter.
    Dotted lines: Results without EVE ($r_N=0$); dashes lines: Results obtained with the free
    (zero baryon density) nucleon radius $r_{N0}$; solid lines: Results obtained with our calculated
    density dependent nucleon radius $r_N$.}
  \label{fig:SNM1}
\end{figure}

\subsection{Symmetric nuclear matter\label{sec:SNM}}
Fig.~\ref{fig:SNM1} shows the results for the binding energy per nucleon and the pressure
in isospin symmetric nuclear matter as functions of the baryon density.
Each figure shows our previous results~\cite{Noro:2023vkx} without EVE (dotted lines), the results where the EVE
are evaluated by using the radius of free nucleons, $r_{N0} = 0.47$ fm as given in the second column of Tab.~\ref{tab:radii}, for all densities (dashed lines), and our full results which include the
density dependence (swelling) of the nucleon radius, which arises via its dependence on the mean
scalar field (solid lines).

We see that at normal nuclear matter density the EVE have only little effect on the EOS. The left panel
of Fig.~\ref{fig:SNM1} shows that the saturation density is shifted from 0.15 fm$^{-3}$ (dotted line) to 0.14 fm$^{-3}$
(solid line), with a corresponding shift of the binding energy per nucleon from $-16\,$MeV to $-14.6\,$MeV.
Although it is possible to readjust model parameters to keep the saturation point unchanged, we
keep the same parameters for clarity, because such small changes would not affect our
main results and conclusions in any way. Also the other results given in Ref.~\cite{Noro:2023vkx} at the saturation
point, like the incompressibility ($K=363$ MeV), the symmetry energy ($a_s=18$ MeV), the effective quark mass
($325$ MeV) and the effective nucleon mass ($756$ MeV), change by less than $5\%$ when the EVE are included.

On the other hand, at densities of about 3 times the normal nuclear matter density, both the energy per nucleon
and the pressure increase appreciably when the EVE are included.
For the energy, about 30$\%$ of the repulsive contributions come from the swelling of the
nucleons in the medium, while for the pressure the swelling effects make up 50$\%$ of the
total EVE. Because typical values of the central baryon densities
in massive neutron stars are about $0.6 \sim 0.8$ fm$^{-3}$, one may naively expect from these figures that the EVE may have a large effect on heavy neutron stars.

\begin{figure}[tbp]
  \subfigure{\centering\includegraphics[width=0.48\columnwidth]{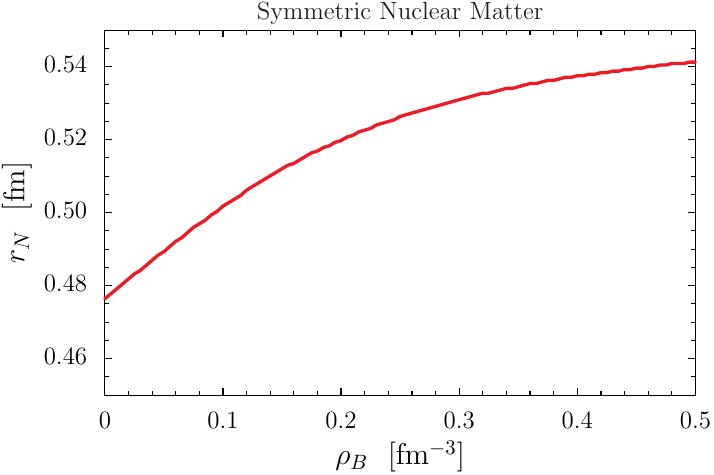}} \hfill
  \subfigure{\centering\includegraphics[width=0.48\columnwidth]{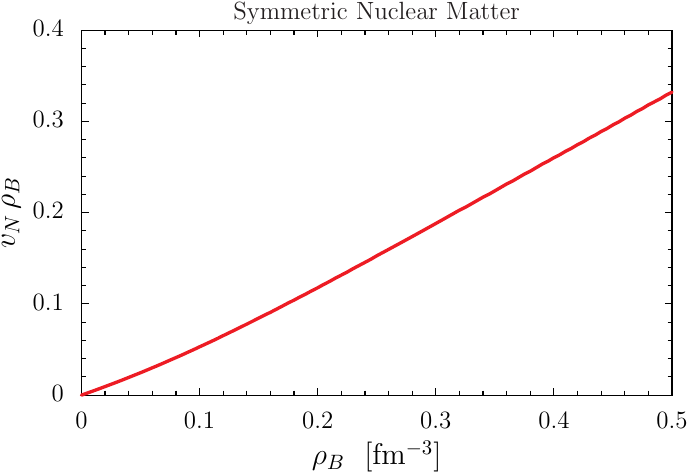}}
  \caption{Nucleon hard core radius ({\it left panel})
    and corresponding volume fractions occupied by the quark cores ({\it right panel}) in
    symmetric nuclear matter as functions of the baryon density.}
  \label{fig:SNM2}
\end{figure}

Fig.~\ref{fig:SNM2} shows the results for the nucleon radius ($r_N$) and the corresponding volume fraction occupied by the quark cores in the medium ($\rho_B \, v_N(\rho_B)$) as functions of the baryon density.
The modest swelling of about $7\%$ at normal densities plays an important role to describe
several nuclear phenomena, as mentioned in Sec.~\ref{sec:intro}. 
At the highest density shown in Fig.~\ref{fig:SNM1}, which is about 3.3 times the normal nuclear matter density,
the swelling of the nucleon radius is about $13\%$ and indicates a kind of plateau, which reflects our
phenomenological implementation of confinement effects via the infra-red cut-off ($\Lambda_{\rm IR}$).
The volume fraction occupied by the quark cores at this density is about 33$\%$, indicating that the physical
picture of the mean field approximation is not unreasonable, but leaving room for effects
from the Pauli principle on the level of quarks, which is an important subject of current research as
mentioned in Sec.~\ref{sec:intro}.

\subsection{Neutron star matter\label{sec:nsm}}
In order to calculate the EOS of neutron star matter, the basic formalism explained in Sec.~\ref{sec:eve} must be supplemented by the conditions for chemical equilibrium and charge neutrality, expressed by 
\begin{align}
  \mu_b - \mu_n + q_b \, \mu_e = \mu_{\mu} - \mu_e = q_i \rho_i = 0  \,.
  \label{eq:cond}
\end{align}
Here the chemical potentials of baryons are given by Eq.~(\ref{eq:mub3}), and for the leptons ($\ell=e^-, \mu^-$) we assume the non-interacting Fermi gas expression $\mu_{\ell} = \sqrt{p_{\ell}^2 + m_{\ell}^2}$. The Fermi momenta $p_i$ of baryons ($i = b$) and leptons ($i=\ell$) are related to their number densities $\rho_i$ by $\rho_i = \frac{p_i^2}{3 \pi^2}$. In Eq.~(\ref{eq:cond}), $q_i$ are the electric charges of baryons and leptons. The numerical calculation is then carried out as follows: The vector fields can be eliminated from the outset in favor of the baryon densities by using Eq.~(\ref{eq:stat}). Then, for a given baryon density $\rho_B$, we solve the set of equations (\ref{eq:cond}) for the number densities of the particles in the system on a grid of scalar fields $\sigma_{\alpha}$, which in turn is determined by minimizing the energy density (\ref{eq:ed}), see Eq.~(\ref{eq:gap}). The solutions of these equations can be used to  calculate the EOS and chemical composition of the system, together with other physical quantities like the effective masses of quarks and baryons, the radii of the baryons and the volume fraction occupied by the quark cores of baryons. 

\begin{figure}[tbp]
  \subfigure{\centering\includegraphics[width=0.48\columnwidth]{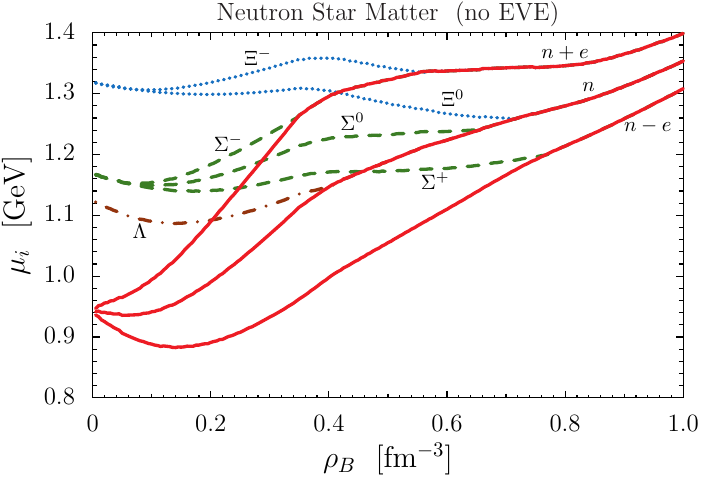}} \hfill
  \subfigure{\centering\includegraphics[width=0.48\columnwidth]{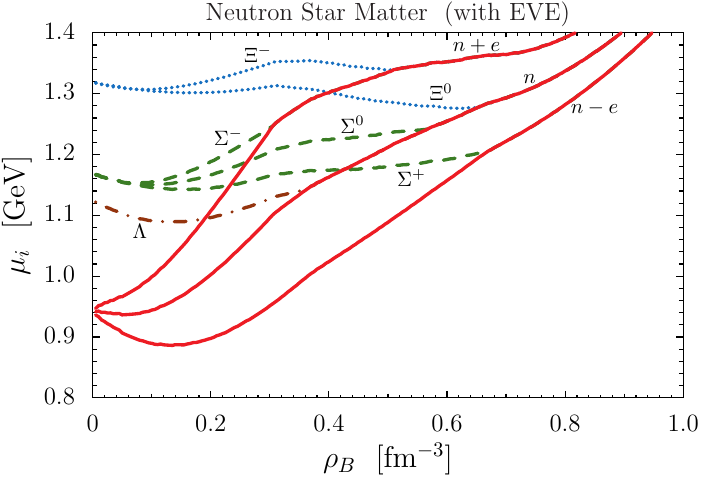}} \\
  \subfigure{\centering\includegraphics[width=0.48\columnwidth]{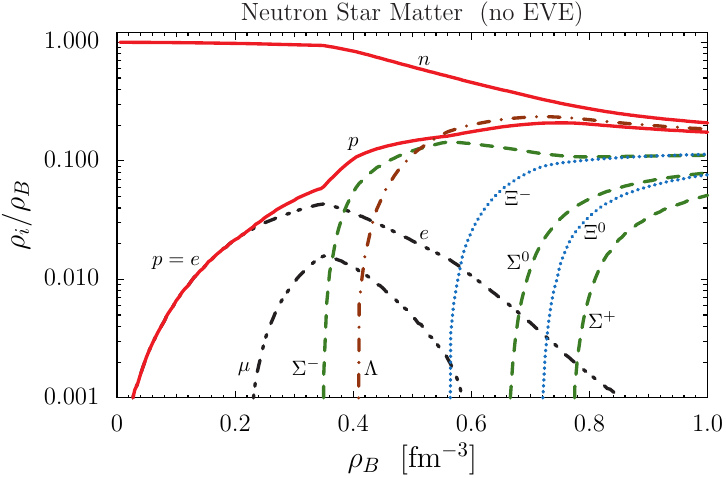}} \hfill
  \subfigure{\centering\includegraphics[width=0.48\columnwidth]{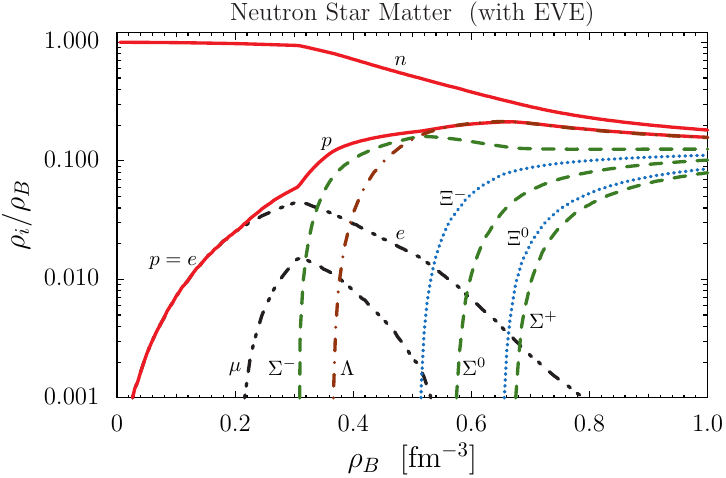}}
  \caption{Chemical potentials of octet baryons ({\it upper panels})
    and ratio of number densities of baryons and leptons to the baryon density ({\it lower panels}) in neutron star matter as functions of the baryon density. The results for $r_b=0$ (no EVE) are shown by the left panels, while those on the right refer to the case of our calculated density dependent baryon radii (with EVE).}
  \label{fig:chem}
\end{figure}

In Fig.~\ref{fig:chem} we compare the results for the chemical potentials and the ratios of number densities of baryons and leptons to the baryon density, for the case where the EVE are neglected (panels on the left) to the case where the EVE are included (panels on the right). The chemical potentials of the hyperons below their threshold densities, where they touch one of the three solid lines representing $\mu_n$ and $\mu_n \pm \mu_e$ from above according to the equilibrium conditions $\mu_b = \mu_n - q_b \mu_e$ (see Eq.~\eqref{eq:cond}), are given by Eq.~(\ref{eq:mub3}) for $\rho_b = p_b = \tilde{p}_b = 0$. They can be visualized as  hyperons immersed in the nuclear medium without forming a nonzero macroscopic number density. 

\begin{figure}[t!]
  \subfigure{\centering\includegraphics[width=0.48\columnwidth]{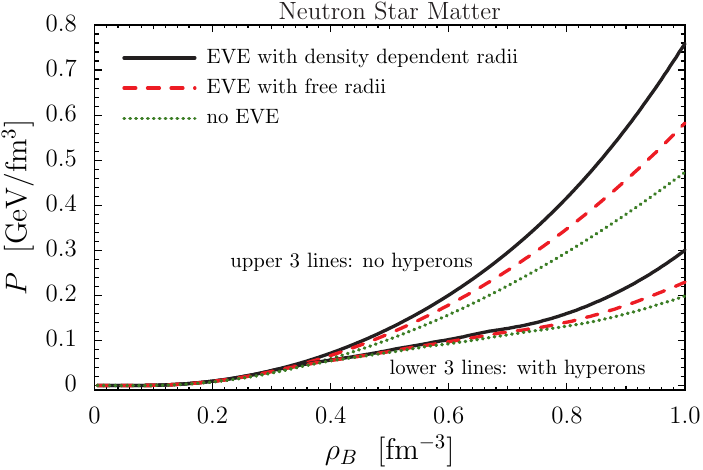}} \hfill
  \subfigure{\centering\includegraphics[width=0.48\columnwidth]{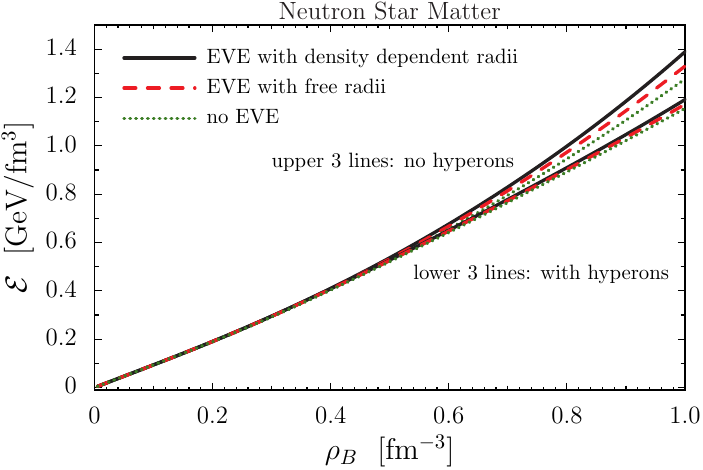}} \\
  \centering\includegraphics[width=0.48\columnwidth]{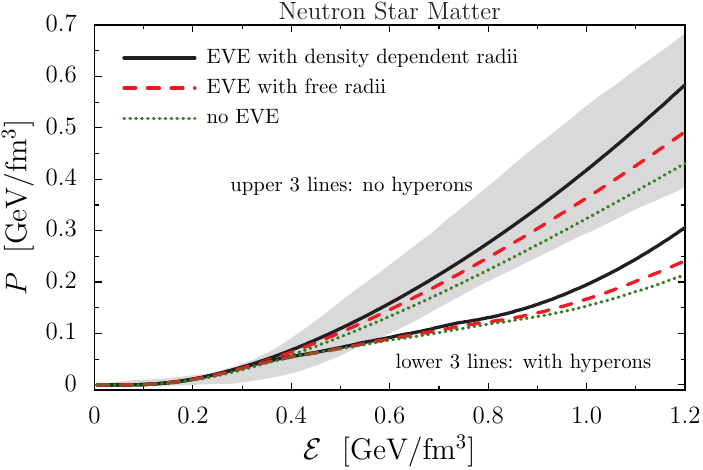}
  \caption{The pressure ({\it upper left panel}) and the energy density ({\it upper right panel})  as functions of the baryon density, and the pressure as a function of energy density ({\it lower panel}), for neutron star matter. In each panel, The upper (lower) three lines show the  results obtained without (with) hyperons, and for each case we show the results without EVE (dotted lines), with EVE calculated with the free baryon radii (dashed lines), and with our calculated density dependent baryon radii (solid lines). The grey area in the lower panel reproduces the $95\%$ posterior credible band shown in Fig.~2 of Ref.~\cite{Brandes:2023bob} from a Bayesian inference analysis of NICER data on neutron stars.}
  \label{fig:nsm1}
\end{figure}

Comparison of the two upper panels of Fig.~\ref{fig:chem} shows that with EVE the  nucleon chemical potentials (solid lines) increase faster with density, partially because of their larger effective Fermi momenta ($\tilde{p}_b$ in Eq.~(\ref{eq:mub3})), and partially because of their volume terms in (\ref{eq:mub3}), while the chemical potentials of the hyperons below their thresholds increase only slightly from their $r_b=0$ values. This is because for $\rho_b=0$ (case of hyperons dispersed in the nuclear medium) the difference between the chemical potentials for $r_b>0$ and $r_b=0$ comes, besides small differences of  effective masses, only from the volume term in Eq.~(\ref{eq:mub3}), which is quite small in the density region of the hyperon onset.\footnote{For example, the value of $P_B$ at $\rho_B=0.4$ fm$^{-3}$, including the EVE, is $P_B = 0.022$ GeV/fm$^{3}$,   which is about $40\%$ of the total pressure at this density, see Fig.~\ref{fig:nsm1}. Then a typical hadron size of $r_b = 0.5$ fm gives a contribution $v_b \, P_B \sim 0.01$ GeV to the chemical potential, which is small on the scale of Fig.~\ref{fig:chem}.} Thus, the crossing between the steeper solid lines and the almost unchanged hyperon lines in Fig.~\ref{fig:chem} is shifted to smaller densities when the EVE are included, and as a result the onset of hyperons occurs in a more narrow window of densities, as shown in the right panels of Fig.~\ref{fig:chem}. Because the presence of hyperons generally tends to soften the EOS, we can therefore expect that the repulsion induced by the EVE will be much less effective in neutron star matter than in nuclear matter.

The pressure and the energy density in neutron star matter are shown in Fig.~\ref{fig:nsm1} for the following six cases: The upper three lines show the results without hyperons (only nucleons and leptons in chemical equilibrium), and the lower three lines show the results including the hyperons. For each case we show the results obtained without EVE ($r_b=0$, dotted lines), with EVE calculated by using the free baryon radii ($r_{b0}$, dashed lines), and with our calculated density dependent baryon radii ($r_b$, solid lines).

Because the energy density is determined mainly by the number densities and the effective masses of the baryons, the EVE on the energy density are generally small. Concerning the pressure, as can be expected from the results of Sec.~\ref{sec:SNM} and the discussions presented above, the EVE are very large in the region of high densities if only nucleons and leptons are considered, but become much smaller when also the hyperons are included. Comparison of the solid lines in the left panel of Fig.~\ref{fig:nsm1} demonstrates the enormous reduction of the pressure by almost a factor of $3$ in the region $\rho_B \simeq 0.8$ fm$^{-3}$. Comparison of the dashed and solid lines shows that for the case without hyperons the swelling of nucleons contributes roughly 50$\%$ to the total EVE for the pressure, similar to the case of symmetric nuclear matter shown in Fig.~\ref{fig:SNM1}, while for the case including the hyperons the swelling effects are dominant.\footnote{The velocity of sound ($c_s = \sqrt{{\rm d}P/{\rm d}{\cal E}}$) does not exceed the velocity of light for the range of densities shown in all figures of this paper. Violations of causality occur in some cases   for baryon densities larger than 1 fm$^{-3}$ when the   EVE are included. At the central baryon densities in our stable neutron stars with maximum mass   (see Sec.~\ref{sec:ns}),   we obtain $c_s = 0.72$ ($0.89$) for the case without hyperons and no EVE (with EVE), and   $c_s=0.39$ ($0.48$) for the case with hyperons and no EVE (with EVE). We note that an upper bound   of $c_s=0.88$ has been reported recently from relativistic hydrodynamics~\cite{Hippert:2024hum}.} The comparison with the $95\%$ posterior credible band~\cite{Brandes:2023bob} in the lower panel of   Fig.~\ref{fig:nsm1} clearly rules out the presence of hyperons in our EOS for $\rho_B \geq 4 \rho_{B0}$, demonstrating the hyperon puzzle in an impressive way.

\begin{figure}[tbp]
  \subfigure{\centering\includegraphics[width=0.48\columnwidth]{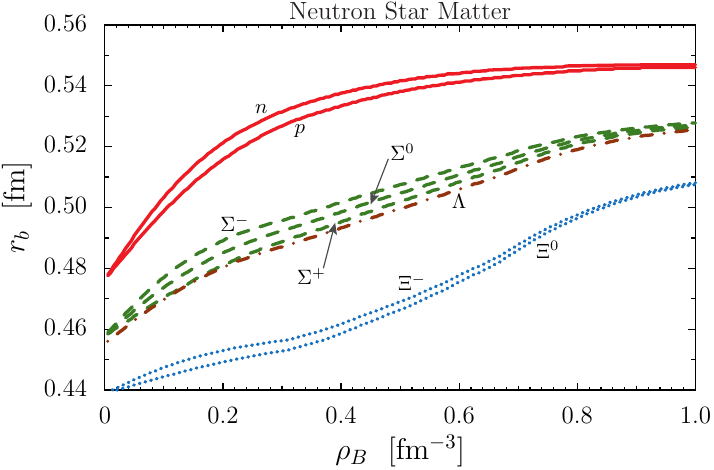}} \hfill
  \subfigure{\centering\includegraphics[width=0.48\columnwidth]{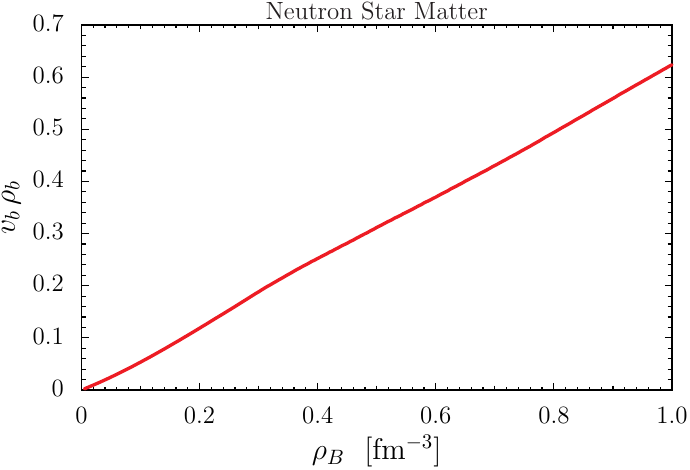}}
  \caption{The calculated radii of octet baryons ({\it left panel}) and the corresponding volume fraction occupied by the quark cores ({\it right panel}) in neutron star matter as functions of the baryon density.}
  \label{fig:nsm2}
\end{figure}

The calculated quark core radii of octet baryons and the corresponding volume fraction occupied
by the quark cores are shown in Fig.~\ref{fig:nsm2}. In the left panel, the density dependence of the radii of hyperons
below their onsets simply reflects the action of the scalar fields produced by other 
baryons with finite density on the quarks inside the hyperon at rest. In the right panel, the volume fraction naturally includes only those baryons which exist in the system with a finite number density.  

From the left panel of Fig. \ref{fig:nsm2} we see that the swelling of baryons in the medium, which is caused by the
decreasing quark masses, is at most ${\sim}20\%$, and\emdash in particular for the nucleons\emdash tends to saturate at high densities.
As mentioned already in Sect.~\ref{sec:SNM}, this reflects our phenomenological implementation of confinement effects via the infrared cut-off
($\Lambda_{\rm IR}$). 
The splittings of radii within the isospin multiplets ($N$, $\Sigma$, $\Xi$) of the baryon octet shown in the left panel of
Fig.~\ref{fig:nsm2} are caused by $M_u > M_d$ in a medium which contains more
$d$ quarks than $u$ quarks. This effect, which gives a small negative contribution to the symmetry energy in
normal nuclear matter, is caused by our scalar isovector part of the effective baryon-nucleon interaction,
which is positive but small at high densities~\cite{Noro:2023vkx}, and which corresponds to the effect of the $\delta$-meson exchange in
baryon-meson theories~\cite{Ulrych:1997es}. The radius of the $\Lambda$ baryon stays below the radii of the $\Sigma$ baryons for all densities.
This can be traced back to the additional attraction in the $\Lambda$ from the
scalar diquark made of $u$ and $d$ quarks (antisymmetric flavor combination $[ud]$), which is not present in the $\Sigma$ baryon,
and which becomes more pronounced at finite density because the mass of this light scalar diquark decreases more rapidly with density than the
mass of the other (heavy scalar and axial vector) diquarks.

The right panel of Fig.~\ref{fig:nsm2} shows that for baryon densities less than ${\sim}0.75$ fm$^{-3}$, which in our calculation is the highest density in stable neutron stars including hyperons, the volume fraction occupied by the quark cores is at most ${\sim}45 \%$. Because this is still well below the closest packing fraction (${\sim}74 \%$ for spheres of similar radii), it may give us some confidence that the overall physical picture of the mean field approximation of 3-quark bound states moving in self consistent scalar and vector fields does not break down at such high densities, but can be used as a starting point for corrections and improvements, taking into account the effects of the Pauli principle on the level of quarks.

\subsection{Neutron stars\label{sec:ns}}
\begin{figure}[tbp]
  \subfigure{\centering\includegraphics[width=0.48\columnwidth]{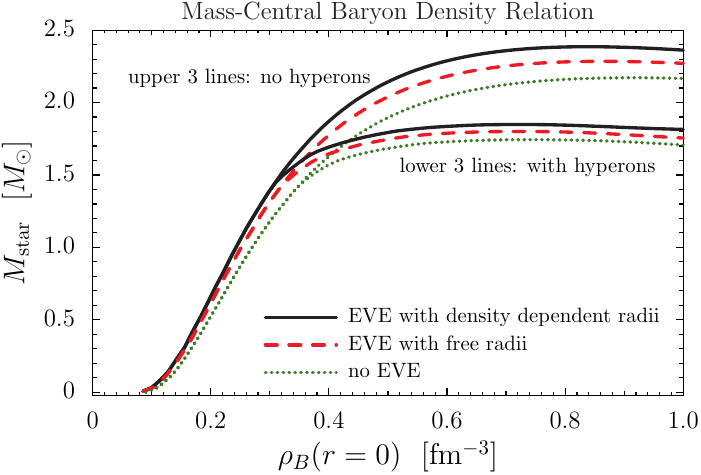}} \hfill
  \subfigure{\centering\includegraphics[width=0.48\columnwidth]{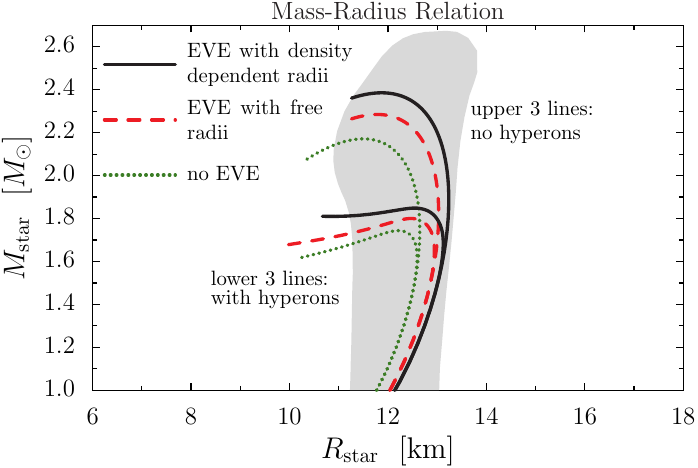}}   
  \caption{Neutron star mass as a function of the central baryon density ({\it left panel}) and the relation between the mass and the radius of the stars ({\it right panel}). In both figures, the upper (lower) three lines show the results obtained without (with) hyperons, and for each case we show the results without EVE (dotted lines), with EVE calculated by using the free baryon radii (dashed lines), and by using our calculated density dependent radii (solid lines). The grey area in the right panel reproduces the $95\%$ posterior credible band shown in Fig.~5 of Ref.~\cite{Brandes:2023bob} from a recent Bayesian inference analysis of NICER data on neutron stars.}
  \label{fig:stars}
\end{figure}

By using our EOS as input to the Tolman-Oppenheimer-Volkoff (TOV) equations~\cite{Tolman:1939jz,Oppenheimer:1939ne}, we can calculate properties of neutron stars.  Our results for the star masses and radii based on the EOS discussed in the previous subsection are collected in Fig.~\ref{fig:stars}.

Both figures demonstrate the reduction of neutron star masses caused by the reduction of pressure in the high density region when the hyperons are included (Fig.~\ref{fig:nsm1}). The two groups of lines shown in Fig.~\ref{fig:stars} clearly separate from each other in the density region of the onset of hyperons.  The repulsive EVE increase the maximum masses of stable stars consisting only of nucleons and leptons by ${\sim}0.3 \, M_{\odot}$ up to $2.39 \,M_{\odot}$, which is larger than the value reported recently for the heaviest observed neutron star~\cite{Romani:2022jhd}.    On the other hand, because of the reasons discussed in the previous subsection, the repulsive EVE become suppressed very much when the hyperons are included, and increase the maximum mass of stable stars by only $0.1\,M_{\odot}$ up to $1.85\,M_{\odot}$. This surprisingly small change is explained by the fact that the EVE, in spite of introducing repulsion into the system, narrow the region of densities for the onset of hyperons, where the direct effect of the finite baryon volumes is too small to balance the gain of energy when a neutron, which gets more kinetic energy as the available volume shrinks, is transformed to a hyperon at rest by the weak interactions. We also mention that the swelling of baryons makes up ${\sim}50\%$ of the total EVE on the maximum star masses in all cases shown in Fig.~\ref{fig:stars}. The comparison with the $95\%$ posterior credible band~\cite{Brandes:2023bob} in the right panel of Fig.~\ref{fig:stars} shows that our stars with masses in the range $ 1.4 \sim 2.0\,M_{\odot}$ tend to have too large radii when the EVE are included. This is because the inner crust of the stars, typically in the density range of ${\sim}10^{-3} \rho_{B0}$ to $\rho_{B0}$, is not included in   our calculation. Its effects can decrease the radius of the stars in this mass region by roughly 1~km~\cite{Rather:2020gja}.

Summarizing, we obtain the following results for the maximum central baryon density which gives stable stars, and the corresponding maximum star mass and radius at maximum mass for the case when the EVE are taken into account with our calculated density dependent baryon radii:
\begin{align}
(\rho_{B}^{\rm max}(r=0), \, M_{\rm star}^{\rm max}, \, R_{\rm star}) 
= ( 0.84 \, {\rm fm}^{-3},  \,\, 2.39 \, M_{\odot}, \,\,  11.84 \, {\rm km})
\nonumber
\end{align}
for the case of no hyperons, and
\begin{align}
(\rho_{B}^{\rm max}(r=0), \, M_{\rm star}^{\rm max}, \, R_{\rm star}) 
= ( 0.73 \, {\rm fm}^{-3},  \,\, 1.85 \, M_{\odot}, \,\,  12.54 \, {\rm km}) 
\nonumber
\end{align}
for the case with hyperons. We therefore conclude that the long standing ``hyperon puzzle'' persists in our relativistic effective quark theory which takes into account the repulsive EVE caused by the finite quark core sizes of the baryons. 

Here we wish to emphasize again the important role of symmetries which has lead us to this conclusion: Our model is based on the chiral and flavor symmetries of the interaction Lagrangians, given in Eq.~(\ref{eq:lagrangian}) for the $\bar{q} q$ channels and in Eq.~(A1) of Ref.~\cite{Noro:2023vkx} for the $qq$ channels, following the well established paths in hadronic physics which lead to many successful low-energy theorems and mass formulas. The essential assumption here is that explicit breaking of these symmetries are allowed to arise only from the current quark masses, and all other breakings must be of dynamical origin. Because chiral symmetric 6-fermi and 8-fermi interactions do not solve the hyperon puzzle~\cite{Noro:2023vkx}, and our aim was to see how far we can go without invoking a phase transition to quark matter, we were left with just 8 model parameters (4 coupling constants, 2 cut-offs, and 2 current quark masses). If these are fixed as explained in Sec.~\ref{sec:model}, we do not have any free parameter to cure disagreements with observations.  If we would allow for symmetry breaking interactions, modifications of the basic model equations etc., there would be many ways to ``resolve'' the problem and get neutron star masses as large as wanted.\footnote{Examples would be to enhance the coupling constant in the vector fields of Eq.~(\ref{eq:fields}) for the $s$ quark, to avoid the decrease of the $s$ quark mass in the medium, to introduce a phenomenological correlation function into the energy density, and many others.} To us, however, it seems more important to respect the basic symmetries observed in nature than to reproduce observations by force.

Based on our results, we expect that the solution of the hyperon puzzle may be difficult to achieve on the level of composite baryons with short range repulsions. Still the possibility of strong 3-body forces involving strangeness on the level of baryons is the subject of  intensive theoretical and experimental investigations~\cite{Tamura:2022sme}, but to us it seems that some kind of strongly interacting quark matter will emerge as the solution, either by invoking a phase transition to color superconducting quark matter~\cite{Lastowiecki:2011hh,Tanimoto:2019tsl}, or by exploring the possibilities opened by recent work on quarkyonic matter~\cite{Fujimoto:2024doc,Kojo:2024ejq,Xia:2024wpz}, which is a new form of high density matter based on the hadron-quark continuity.

\section{Summary\label{sec:summary}}
Strongly interacting baryonic systems are fascinating objects of current research in hadronic physics, because they relate the basic symmetries of microscopic nuclear systems to macroscopic astrophysical objects. In our work, we used the NJL model as an effective quark theory of QCD to describe constituent quarks, relativistic bound states forming hadrons, and the equations of state for nuclear and neutron star matter. This model has the basic chiral and flavor symmetries of QCD, which must be respected by all models for the hadronic interactions. Our basic assumption was that the current quark masses are the only sources of explicit symmetry breakings, while all other mechanisms of symmetry breaking must be of dynamical origin. This basic assumption leads to very strong restrictions on the model parameters, which have to be respected irrespective of agreement or disagreement with observations.

The new point of our work was a consistent description of excluded volume effects (EVE) in the NJL model by identifying the quark core radii of each baryon in the system with the hard distance scale at which the relative wave function of interacting baryons becomes strongly suppressed. For this purpose, we reformulated the EOS including EVE such that each baryon can have its own quark core size, which we calculated consistently within our model. We have shown that this description of EVE, which does not introduce any new parameters, satisfies the requirements of thermodynamic consistency in general, and of causality up to densities more than 6 times the normal nuclear matter density, which exceeds the range which is relevant for the cores of heavy neutron stars. By solving the 3 independent gap equations for the $u$, $d$, and $s$ quarks, we were able to dynamically include the effects of in-medium hadron swelling, which are known to be important for the case of the nucleon, on the EOS at high baryon densities. Our main motivation to re-examine the role of EVE was to investigate their effects on the long standing hyperon puzzle.

Our findings can be summarized as follows: For symmetric nuclear matter the mechanism of volume exclusion by the swelling quark cores leaves the saturation properties almost unchanged, but at about 3 times the normal density it increases both the energy per nucleon and the pressure by about 30$\%$, while increasing the quark core size of the nucleons modestly by 10$\%$ and keeping the volume fraction occupied by the quark cores below 30$\%$. These results reflect our phenomenological implementation of confinement effects into the model. A stable neutron star made only of nucleons and leptons can then be as heavy as ${\sim}2.4$ solar masses, without any violation of causality. However, the direct repulsive effects coming from the quark core sizes are too small to balance the increased energy gain from conversion of fast nucleons into hyperons at rest by the weak interactions. This effect decreases the density window where hyperons appear by ${\sim}15\%$, which is enough to cancel most of the direct repulsive EVE on the pressure at high densities and the maximum mass of stable neutron stars. The result that we can get stable stars of only up to $1.85$ solar masses if the hyperons and EVE are included together leads us to expect that the hyperon puzzle may be difficult to solve on the level of composite baryons with short range repulsions.

Besides further investigations on strong 3-body forces involving strange baryons, there are a number of ways to proceed by exploiting explicit quark degrees of freedom. One natural way might be to connect our NJL EOS with a color superconducting 3-flavor quark EOS calculated in the same model by the usual Maxwell or Gibbs constructions. However, there are also other possibilities as reported recently. For example, in Ref.~\cite{Fukushima:2020cmk} it was argued that the hard cores should be replaced by continuous distributions, i.e., by the radial variable of the density, energy and pressure profiles of a single baryon. In this way it was observed that the resulting EOS, based on a closest packing of nucleons with variable radial size variables, becomes similar to empirical EOSs from other approaches. Another line proposed in the same paper is based on quantum percolation and hadron-quark continuity, where the quark wave functions become delocalized without a phase transition. It was pointed out, and supported by recent work based on schematic forms for the quark momentum distribution in baryons~\cite{Fujimoto:2024doc,Kojo:2024ejq}, that this mechanism could lead to a solution of the hyperon puzzle. The fascination of studying strongly interacting matter at high baryon densities is continuing, and will increase our knowledge about the basic building blocks of nature.

\acknowledgments{This work was supported by the U.S. Department of Energy, Office of Science, Office of Nuclear Physics, contract no. DE-AC02-06CH11357.}

\conflictsofinterest{The authors declare no conflicts of interest.}

\appendixtitles{yes}
\appendixstart

\appendix

\section[\appendixname~thesection]{EFFECTIVE INTERACTION BETWEEN BARYONS\label{app:interaction}}

By using Landau's Fermi liquid theory~\cite{Negele:1988aa,Shankar:1993pf}, the $\ell=0$ effective interaction
between two baryons ($b$ and $b'$) at their
respective Fermi surfaces, denoted by $f_{0, b b'}$, can be obtained from Eq.~(\ref{eq:mub3}) of the main text
as follows:
\begin{align}
  \frac{{\rm d} \mu_b}{{\rm d} \rho_{b'}} = \frac{\delta_{b b'}}{(1 - v \cdot \rho) N_b} + f_{0, b b'}  \,,
  \label{eq:landau}
\end{align}
where $N_b= \frac{E_b(\tilde{p}_b) \, \tilde{p}_b}{\pi^2}$ is the density of states per volume of baryon $b$
at its Fermi surface, and the first term in (\ref{eq:landau}) refers to the volume available for the
Fermi motion, $\tilde{V} = V(1 - v \cdot \rho)$.

To calculate the derivative from (\ref{eq:mub3}), it is convenient to separate the term depending
explicitly on the density from the terms depending implicitly on the density via the mean fields
$\sigma_{\alpha}$ and $\omega_{\alpha}$:
\begin{align}
  \frac{{\rm d} \mu_b}{{\rm d} \rho_{b'}}  = \frac{{\partial} \mu_b}{{\partial} \rho_{b'}}
  + \frac{{\partial} \mu_b}{{\partial} \sigma_{\alpha}} \frac{\partial \sigma_{\alpha}}{\partial \rho_{b'}}
  + \frac{{\partial} \mu_b}{{\partial} \omega_{\alpha}} \frac{\partial \omega_{\alpha}}{\partial \rho_{b'}} \,.
  \label{eq:split}
\end{align}
The explicit density dependence in (\ref{eq:mub3}) comes from the effective Fermi momentum $\tilde{p}_b$ in $E_b(\tilde{p}_b)$ and the baryon pressure $P_B^{(0)}(\tilde{\rho})$, while the baryon mass $M_b$, the vector potential contribution, and the quark core
volume $v_b$ can be considered as functions of the mean fields. The evaluation of the first term
in (\ref{eq:split}) is straight forward, and for the other terms is is convenient to note that
the energy density is minimized w.r.t. to the mean fields at each set of baryon densities separately, i.e.,
\begin{align}
\frac{\partial}{\partial \rho_{b}} \left( \frac{\partial {\cal E}}{\partial \sigma_{\alpha}} \right)
&= 0 = \frac{\partial {\mu_{b}}}{\partial \sigma_{\alpha}} 
+ \frac{\partial^2 {\cal E}}{\partial \sigma_{\alpha} \partial \sigma_{\beta}} \, 
\frac{\partial \sigma_{\alpha}}{\partial \rho_{b}} \,, 
&
\frac{\partial}{\partial \rho_{b}} \left( \frac{\partial {\cal E}}{\partial \omega_{\alpha}} \right)
&= 0 = \frac{\partial {\mu_{b}}}{\partial \omega_{\alpha}} 
+ \frac{\partial^2 {\cal E}}{\partial \omega_{\alpha} \partial \omega_{\beta}} \, 
\frac{\partial \omega_{\beta}}{\partial \rho_{b}} \,,
\label{eq:relations}
\end{align}
where we used the fact that there is no mixing between scalar and vector mean fields in our energy
density (\ref{eq:ed}).

\begin{figure}[tbp]
\centering\includegraphics[width=0.3\columnwidth]{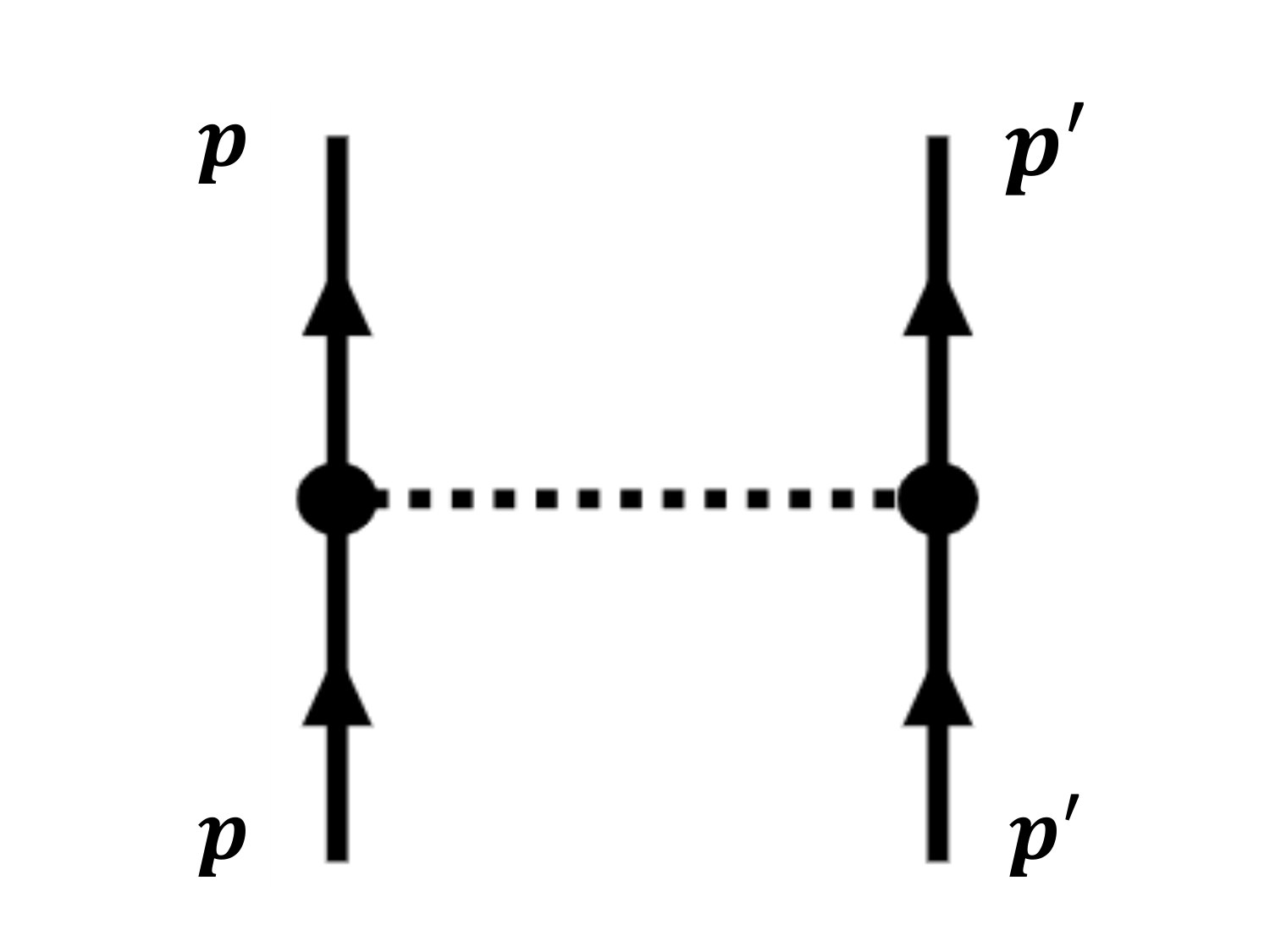}
\caption{Graphical representation of the meson exchange contribution to the effective baryon-baryon interaction, Eq.~(\ref{eq:meson}). The magnitudes of the 3-momenta indicated in this figure
are the Fermi momenta of the baryons $b$ and $b'$, i.e., $|\vec{p}| = p_b$ and $|\vec{p'}| = p_{b'}$.}
    \label{fig:interaction}
\end{figure}

We obtain the following result for the effective interaction:
\begin{align}
  f_{0, b b'} &= \frac{1}{(1 - v \cdot \rho)^2} \left[ \frac{v_b \, \rho_{b'}}{N_{b'}}
  + \frac{v_{b'} \, \rho_{b}}{N_{b}} + \frac{(v_b \, \rho_{b_1}) (\rho_{b_1} \, v_{b'})}
  {(1 - v \cdot \rho) N_{b_1}} \right] \label{eq:ex} \\
   & -\frac{\partial {\mu_{b}}}{\partial \sigma_{\alpha}} \, \left(S^{-1} \right)_{\alpha \beta}
     \, \frac{\partial {\mu_{b'}}}{\partial \sigma_{\beta}}
   -\frac{\partial {\mu_{b}}}{\partial \omega_{\alpha}} \, \left(V^{-1} \right)_{\alpha \beta}
     \, \frac{\partial {\mu_{b'}}}{\partial \omega_{\beta}} \,.
     \label{eq:meson}
\end{align}
Here we defined the generalized scalar and vector meson propagators for zero momentum by the
inverse of the $3 \times 3$ flavor matrices
\begin{align}
S_{\alpha \beta} &\equiv \frac{\partial^2 {\cal E}}{\partial \sigma_{\alpha} \partial \sigma_{\beta}}
\,,  &  V_{\alpha \beta} &\equiv \frac{\partial^2 {\cal E}}{\partial \omega_{\alpha} \partial \omega_{\beta}} \,.
\label{eq:propagators}
\end{align}
The term (\ref{eq:ex}) is the repulsive interaction arising directly from the finite quark core
sizes. The two terms in (\ref{eq:meson}) correspond to the direct terms arising from the exchange of
neutral scalar and vector mesons, as illustrated by Fig.~\ref{fig:interaction}.
The vector meson exchange term in Eq.~(\ref{eq:meson}) is
the same as without EVE, and is simply given by $4 G_v \, n_{\alpha/b}\, n_{\alpha/b'}$. The scalar meson exchange
term is more complicated than the form without EVE given in Ref.~\cite{Noro:2023vkx}). The square of the coupling
constant between a scalar meson made of ($q_{\alpha} \overline{q}_{\alpha}$) and the baryon $b$ is expressed by
\begin{align}
  \frac{\partial {\mu_{b}}}{\partial \sigma_{\alpha}}
  = \frac{M_{b}}{E_b(\tilde{p}_b)} \, \frac{\partial M_b}{\partial \sigma_{\alpha}}
  + \frac{\partial \left(v_b P_B^{(0)}(\tilde{\rho})\right)}{\partial \sigma_{\alpha}} \,,
  \label{eq:coupling}
\end{align}
where the first term has a familiar form~\cite{Bentz:2001vc}, but the second term could only be assessed numerically
because of the dependence of $v_b$ on the scalar fields. For the same reason, also the scalar meson propagator
is more complicated because of the dependence of the energy density (\ref{eq:ed}) on the quark core radii.
The $\bar{u} u$, $\bar{d} d$ and $\bar{s} s$ components of the exchanged scalar meson could be disentangled by an orthogonal transformation to diagonalize $S$ at fixed baryon density. Although further numerical analyses may be
interesting, we do not go into more details in this work.

\begin{adjustwidth}{-\extralength}{0cm}
\reftitle{References}
\bibliography{bib}
\end{adjustwidth}

\end{document}